\documentclass[runningheads]{llncs}
\usepackage{cite}
\usepackage{amsmath,amssymb,amsfonts}
\usepackage{graphicx}
\usepackage{xcolor}
\usepackage{subfig}
\usepackage{multirow}
\usepackage{booktabs}
\usepackage{tcolorbox}
\usepackage{stfloats}
\usepackage{soul}
\usepackage[ruled,linesnumbered]{algorithm2e}

\newcommand\todo[1]{\textcolor{red}{}}

\newcommand{\nmodels}{\not\mathrel|\joinrel=}

\newcommand\edit[1]{\textcolor{orange}{}}

\newcommand{\eventually}[0]{\Diamond}
\newcommand{\globally}[0]{\Box}
\newcommand{\estrobustness}[0]{\widehat{\mathbf{\Delta}}}

\begin{document}

\title{Tolerance of Reinforcement Learning Controllers against Deviations in Cyber Physical Systems}
\author{Changjian Zhang\inst{1}
\thanks{Both authors contributed equally to this research.} 
\and Parv Kapoor\inst{1}
\protect\footnotemark[\value{footnote}] 
\and
Eunsuk Kang\inst{1} \and
R\^omulo Meira-G\'oes \inst{2} \and
David Garlan\inst{1}\and Akila Ganlath \inst{3} \and Shatadal Mishra \inst{3} \and Nejib Ammar \inst{3}}

\institute{Carnegie Mellon University, Pittsburgh PA, USA \and
The Pennsylvania State University, 
State College PA, USA \and
Toyota InfoTech Labs
 Mountain View, CA USA }
\maketitle

\thispagestyle{plain}
\pagestyle{plain}

\begin{abstract}
Cyber-physical systems (CPS) with reinforcement learning (RL)-based controllers are increasingly being deployed in complex physical environments such as autonomous vehicles, the Internet-of-Things (IoT), and smart cities. An important property of a CPS is \emph{tolerance}; i.e., its ability to function safely under possible disturbances and uncertainties in the actual operation. In this paper, we introduce a new, expressive notion of tolerance that describes how well a controller is capable of satisfying a desired system requirement, specified using \emph{Signal Temporal Logic (STL)}, under possible \emph{deviations} in the system. Based on this definition, we propose a novel analysis problem, called the \emph{tolerance falsification problem}, which involves finding small deviations that result in a violation of the given requirement. We present a novel, two-layer simulation-based analysis framework and a novel search heuristic for finding small tolerance violations. To evaluate our approach, we construct a set of benchmark problems where system parameters can be configured to represent different types of uncertainties and disturbances in the system. Our evaluation shows that our falsification approach and heuristic can effectively find small tolerance violations.
\end{abstract}

\section{Introduction}












The \emph{tolerance} of a CPS characterizes the ability of an engineered system to function correctly in the presence of uncertainties. Modern \emph{cyber-physical systems (CPS)} operate in dynamic and uncertain environments, such as autonomous vehicles, medical devices, the Internet of Things (IoT), and smart cities. The mission-critical and safety-critical nature of CPS accentuate the need to provide a high level of tolerance against uncertainties, as a failure to do so could result in severe consequences, from safety hazards to economic losses.

As CPS grow in complexity and scale, \emph{reinforcement learning (RL)} techniques are gaining popularity for learning CPS controllers. In general, these controllers perceive the state of the CPS and take an action that maximizes the \emph{long-term utility}. The utility is captured through reward functions designed by engineers. An RL controller is trained via a trial-and-error process where an agent takes actions in a simulator of the CPS and uses the simulated results of the actions
to discover an optimal control strategy. Hence, the fidelity of the simulator plays a big role in the effectiveness of a trained controller.
Often, there are reality gaps between the actual deployed environment and the simulator due to approximation and under-modeling of physical phenomena, which makes controllers trained in simulations perform poorly in the real world\cite{Collins2018QuantifyingTR}. This performance degradation can also manifest as unsafe system behaviors in the actual environment. 

To make an RL controller tolerant of possible errors due to these reality gaps, existing works often focus on the training stage, such as robust RL \cite{Moos-RobustRL,xu2022trustworthy} and domain randomization \cite{Peng2018-Sim-to-Real,Sadeghi2017cad2rl,Tobin2017domain}. They investigate the problem of training a controller that is capable of maintaining desired system behavior in the presence of possible \emph{system deviations}---environmental uncertainties, observation or actuation errors, disturbances, and modeling errors.
However, these methods are  limited in how desired system behaviors are expressed. In RL, the desired behavior is often expressed using a reward function \cite{Moos-RobustRL,xu2022trustworthy}; it is well-known that encoding a high-level system requirement using a reward function is a challenging task that requires a significant amount of domain expertise and manual effort via reward shaping \cite{10.5555/645528.657613, Booth_Knox_Shah_Niekum_Stone_Allievi_2023}. Additionally, certain requirements cannot be directly encoded as rewards, especially those that capture time-varying behavior (e.g., ``the vehicle must come to a stop in the next 3 seconds").

Due to the limitation in reward functions and the data-driven nature of RL, these training-oriented methods in general do not provide formal guarantees about tolerance. Also, there is a lack of focus on post-training analysis for the tolerance of RL controllers, especially in the sense of maintaining a desired, complex system specification. Moreover, a formal definition of tolerance for RL controllers with respect to system behavior (beyond rewards) is also missing.

To fill the missing gap in post-training tolerance analysis of RL controllers, we propose a new notion of tolerance based on specifications in \emph{Signal Temporal Logic (STL)}~\cite{stl}. Our definition assumes a \emph{parametric} representation of a system, where \emph{system parameters} capture the dynamics of the system (e.g., acceleration of a nearby vehicle) that are affected by system deviations (e.g., sensor errors). A system is initially assigned a set of nominal parameters that describe its expected dynamics. Then, a change in parameters, denoted by $\delta$, corresponds to a deviation that may occur. Finally, a controller is said to be \emph{tolerable} against certain deviations with respect to a STL specification if and only if the controller is capable of \emph{satisfying} the specification even under those deviations.
Based on this tolerance definition, we propose a new type of analysis problem called the \emph{tolerance falsification}. The goal is to find deviations in system parameters that result in a violation of the desired system specification. Specifically, we argue that identifying a violation closer to the nominal system parameters would be more valuable, since such a violation is more likely to occur in practice. Intuitively, our system needs to tolerate these deviations before addressing the ones that are further away from the nominal set. These identified violations could  be used to retrain the controller for improved tolerance, or to build a run-time monitor to detect when the system  deviates into an unsafe region. 

In addition, we propose a novel simulation-based framework where the tolerance falsification problem is formulated as a two-layer optimization problem. In the lower layer, for a given system deviation $\delta$ (representing a particular system dynamics), an optimization-based method is used to find a falsifying signal; i.e., a sequence of system states that results in a violation of the given STL specification. In the upper layer, the space of possible deviations is explored to find small deviations that result in a specification violation, repeatedly invoking the lower-layer falsification.
The results generated from the lower layer guide the upper-layer search towards small violating deviations.
Furthermore, we present a novel heuristic that leverages the differences between the trajectories from the normative and deviated environments, captured via cosine distances, to improve the effectiveness of the upper layer search algorithm.

To evaluate the effectiveness of our falsification approach, we have constructed a set of benchmark case studies. In particular, these benchmark systems are configurable with system parameters to generate a range of systems with different behaviors due to the parameters' impact on how the system evolves.
Our evaluation shows that our approach can be used to effectively find small deviations that cause a specification violation in these systems.


This paper makes the following contributions:
\begin{itemize}
    \item We present a novel, formal definition of tolerance for RL controllers (Sec.~\ref{sec:robustness}), and a new analysis problem named \emph{tolerance falsification problem} (Sec.~\ref{sec:falsification}).
    \item We propose a two-layer optimization-based method and a novel search heuristic for finding small violating deviations (Sec.~\ref{sec:framework}).
    \item We present an RL tolerance analysis benchmark and evaluate the effectiveness of our approach through experimental results on it (Sec.~\ref{sec:evaluation}).
\end{itemize}


\section{Preliminaries}\label{sec:background}

\subsubsection{Markov Decision Process}
We model the systems under study as discrete-time stochastic systems in \emph{Markov Decision Processes} (MDPs)
\cite{handbook-probabilistic}. An MDP is a tuple $\mathbf{M} = \langle S, A, T, I, R\rangle$, where $S \subseteq \mathbb{R}^n$ is the set of states, $A \subseteq \mathbb{R}^m$ is the set of actions (e.g., control inputs), $T: S \times A \times S \to [0, 1]$ is the transition function where $T(s, a, s')$ represents the probability from state $s$ to $s'$ by action $a$ and $\forall s \in S, a \in A: \sum_{s' \in S} T(s, a, s') = 1$, $I: S \to [0, 1]$ is the initial state distribution, and $R: S \to \mathbb{R}$ is the reward function mapping states to a real value. As is often the case for real-world systems, we assume that the transition function is unknown.

We consider black-box deterministic control policies for a system. Formally, a policy $\pi: S \to A$ for an MDP maps states to actions. Reinforcement learning (RL) \cite{sutton2018reinforcement} is the process of learning an optimal policy $\pi^*$ that maximizes the cumulative discounted reward for this MDP. Additionally, a trajectory $\sigma$ of an MDP given an initial state $s_0 \sim I$ and a policy $\pi$ is defined accordingly as
$\sigma = (s_0 \xrightarrow{a_0} s_1 \ldots s_i \xrightarrow{a_i} s_{i+1} \ldots)$
where $a_i = \pi(s_i)$ and $s_{i+1} \sim T(s_i, a_i)$. Finally, we use $\mathcal{L}(\mathbf{M}||\mathbf{\pi})$ to represent the behavior of the controlled system, i.e., it is the set of all trajectories of a system $\mathbf{M}$ under the control of $\mathbf{\pi}$.

\subsubsection{Signal Temporal Logic}\label{sec:stl}
A signal $\mathbf{s}$ is a function $\mathbf{s}: T \to D$ that maps a time domain $T \subseteq \mathbb{R}_{\geq 0}$ to a $k$ real-value space $D \subseteq \mathbb{R}^k$, where $\mathbf{s}(t) = (v_1, \ldots, v_k)$ represents the value of the signal at time $t$. Then, an STL formula is defined as:
$$\phi := \mu ~|~ \neg \phi ~|~ \phi \land \psi ~|~ \phi \lor \psi ~|~ \phi ~\mathcal{U}_{[a,b]}~\psi$$
where $\mu$ is a predicate of the signal $\mathbf{s}$ at time $t$ in the form of $\mu \equiv \mu(\mathbf{s}(t)) > 0$ and $[a, b]$ is the time interval (or simply $I$). The \emph{until} operator $\mathcal{U}$ defines that $\phi$ must be true until $\psi$ becomes true within a time interval $[a, b]$. Two other operators can be derived from \emph{until}: \emph{eventually} ($\eventually_{[a,b]}~\phi := \top~\mathcal{U}_{[a,b]}~\phi$) and \emph{always} ($\globally_{[a,b]}~\phi := \neg\eventually_{[a,b]}~\neg\phi$).

The satisfaction of an STL formula can be measured in a quantitative way as a real-valued function $\rho(\phi, \mathbf{s}, t)$ (also known as the STL \emph{robustness} value), which represents the difference between the actual signal value and the expected one \cite{stl}. For example, given a formula $\phi \equiv \mathbf{s}(t) - 3 > 0$, if $\mathbf{s} = 5$ at time $t$, then the satisfaction of $\phi$ can be evaluated by $\rho(\phi, \mathbf{s}, t) = \mathbf{s}(t) - 3 = 2$. The definition of $\rho$ is as follows ($\rho$ for the other operators can be formulated from these):
\begin{align*}
    & \rho(\mu, \mathbf{s}, t) = \mu(\mathbf{s}(t)) \qquad\qquad \rho(\neg \phi, \mathbf{s}, t) = - \rho(\phi, \mathbf{s}, t) \\
    & \rho(\phi \land \psi, \mathbf{s}, t) = \min \{ \rho(\phi, \mathbf{s}, t), \rho(\psi, \mathbf{s}, t) \} \\
    & \rho(\phi~\mathcal{U}_{I}~\psi, \mathbf{s}, t) = \sup_{t_1 \in I + t} \min \{ \rho(\psi, \mathbf{s}, t_1), \inf_{t_2 \in [t, t_1]} \rho(\phi, \mathbf{s}, t_2) \}
\end{align*}






\section{Motivating Example}\label{sec:motivation}
\begin{figure}[!t]
\centering
\includegraphics[width=0.85\linewidth]{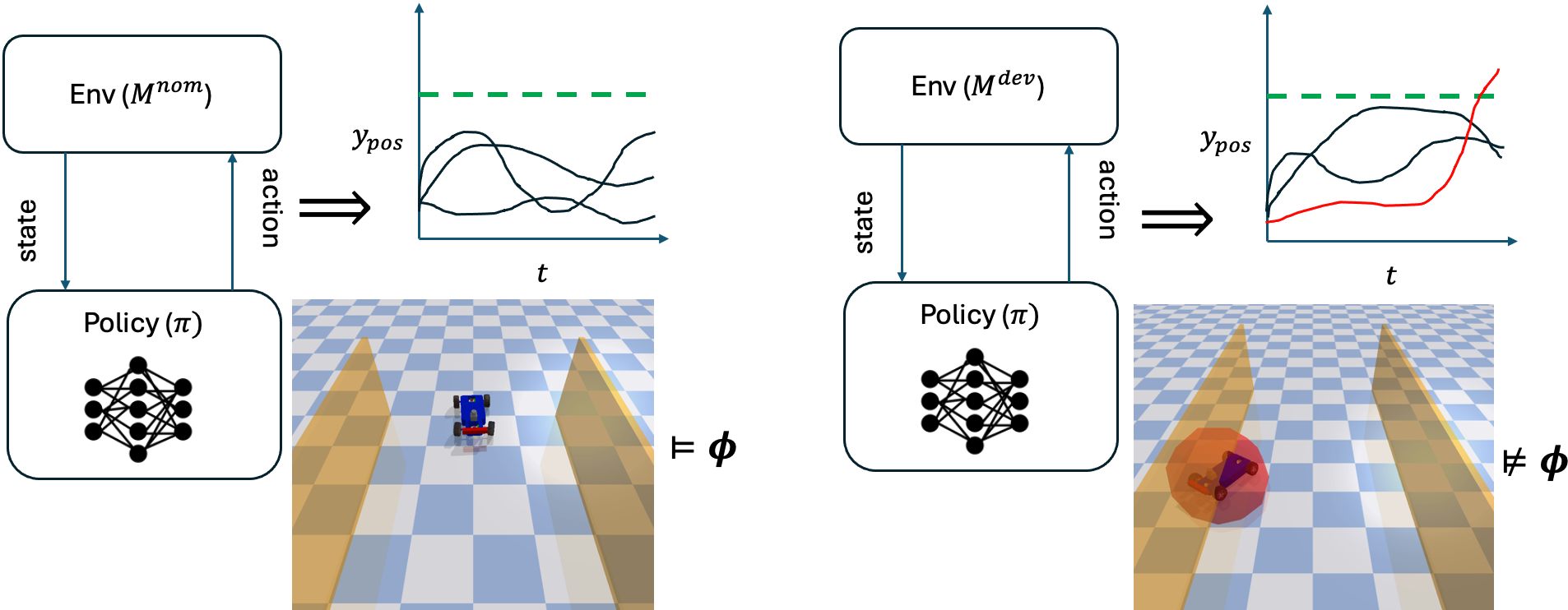}
\caption{\footnotesize{Behavior of the CarRun system under different system parameters.
In the norminal condition (left), $y_{pos}$ in all trajectories is below the threshold (green line) and thus the system is safe. However, in the deviated condition (right), there exists a trajectory where $y_{pos}$ exceeds the threshold and hence the safety requirement is violated.}}
\label{fig:cart-pole-example}
\vspace{-15pt}
\end{figure}

We use an RL system which is required to satisfy a safety specification to illustrate our tolerance definition and analysis. 
Consider the \emph{CarRun} safe RL system implemented in bullet-safety-gym\footnote{https://github.com/SvenGronauer/Bullet-Safety-Gym}, depicted in Figure~\ref{fig:cart-pole-example}.
The CarRun system has a four-wheeled agent based on \textit{MIT Racecar}\footnote{https://github.com/mit-racecar} placed between two safety boundaries. The safety boundaries are non-physical bodies that can be breached without causing a collision. The objective is to go through the avenue between the boundaries without penetrating them. The agent velocity also needs to be maintained below a user-defined threshold. 
Formally, it can be specified by an STL invariant: $\Box (|y_{pos}| < C_1 \land |v| < C_2)$, where $C_1$ and $C_2$ are the constant thresholds for the y coordinate and the velocity, respectively.

Given the CarRun system, we can
train an RL controller such that the car agent satisfies the safety specification above using methods from safe RL \cite{10.5555/2789272.2886795} \cite{Gu2022ARO}. 
However, to transfer this ``safe'' controller to the real world, we need to account for the reality gap between the simulator and the deployed environment. This reality gap might arise due to inaccurate modeling of contact surfaces, actuator errors, and incorrect physical parameter configuration (e.g., friction and mass). These reality gaps can lead to the agent violating the safety specification in the real world, despite satisfying them in simulation. Additionally, since the RL controllers are black-box neural networks, it is extremely hard to capture their concrete behaviors. The difficulty in reasoning about the controller's behaviors coupled with the stochasticity of the system leads to a challenging analysis problem of understanding their tolerance ability.
This has long been one of the key drawbacks that limit the application of these controllers in the real world \cite{Tobin2017domain,yu2018policy}.


Since it can be challenging to quantitatively measure these reality gaps, we take a parametric approach. We approximate the reality gap between the simulator and the deployed environment quantitatively using deviations as parameters. For example, we model the CarRun system as being \emph{parametric} with two controllable \emph{system parameters}, $tm$ (turn multiplier, a factor for the steering control) and $sm$ (speed multiplier, a factor for the speed control).
These parameters govern the impact of the action provided by the controller, e.g., a larger $sm$ will result in more aggressive accelerations. The intuition behind these deviations is to account for actuation issues that arise while deploying agents in the real world.
Figure~\ref{fig:cart-pole-example} shows the behavior of CarRun under different system parameters. 
In Figure~\ref{fig:cart-pole-example}(a), the agent is deployed in the nominal condition with default system parameters.
In this scenario, the controller successfully manages to drive the Car agent through the avenue and also maintains a safe velocity, i.e., the safety specification is satisfied.
In Figure~\ref{fig:cart-pole-example}(b), we show the same controller deployed under a deviated CarRun environment with different turn and speed multipliers. 
In this scenario, the controller makes the car behave erratically, which eventually makes the car cross the safety boundary, i.e., the safety specification is violated.

This example highlights the brittleness of these controllers concerning safety specifications and the need for stakeholders to address 
pre-deployment questions like:
\emph{What are the possible deviations that these RL controllers can tolerate?}
More specifically, \emph{how much change in the system parameters can the controller tolerate before it begins to violate the given safety specification?} We formulate this question as a type of analysis problem called \emph{tolerance falsification},
where the goal is to find deviations in system parameters (e.g., the changes in the turn and speed multiplier of CarRun) where the deviated system violates the given specification. This analysis problem is challenging due to the stochastic, black-box nature of the system as well as the opacity of NN-based RL controllers. 

Additionally, a notion of ``quality of solution''
while searching for system parameters is necessary to factor in the practical assumptions about the operating context of this system. For example, deviations that are closer to the nominal parameters
are more likely to occur in practice and hence need to be prioritized when analyzing. This helps avoid impractically large deviation values that might cause a violation but offers little insight to system designers.
Thus, our falsification process attempts to find violations with \emph{small} deviations; i.e., minimal parameter changes that introduce a risk of specification violation into the system.
The output of this analysis (i.e., violations) can help the engineer identify RL-based controller brittleness and can be used to redesign or retrain the controller to improve its tolerance.

\section{Tolerance Definition}\label{sec:robustness}

\subsection{Definition of Specification-Based Tolerance}
In this work, we use STL to specify the desired properties of a system, and system parameters to capture the deviations in system dynamics. Parameters can represent a variety of deviations such as environmental disturbances (e.g., wind or turbulence),  internal deviations (e.g., mass variation of a vehicle), observation errors (e.g., sensor errors), or actuation errors (e.g., errors in steering control). Then, to capture systems with such diverse dynamics using parameters, we leverage the notion of \emph{parametric control systems}\cite{Bhattacharyya:1995,weinmann:2012uncertain}.

A \emph{parametric} discrete-time stochastic system $\mathbf{M}^\Delta$ defines a set of systems such that $\Delta \subseteq \mathbb{R}^k$ represents the parameter domain, and for any $\delta \in \Delta$, an instance of a parametric system $\mathbf{M}^\delta$ is an MDP $\mathbf{M}^\delta = \langle S, A, T^\delta, I^\delta, R \rangle$,
where the initial state distribution $I^{\delta}$ and the state transition distributions $T^{\delta}$ are both defined by the parameter $\delta$.
Parameter $\delta$ represents a deviation to a system and $\Delta$ represents the domain of all deviations of interest. In addition, we use $\delta_0 \in \Delta$ to represent the zero-deviation point, i.e., the parameter under which the system $\mathbf{M}^{\delta_0}$ exhibits the expected, normative behavior. Then, we define a system as being tolerable against a certain deviation as follows:
\begin{definition}\label{def:robust-system}
For a system $\mathbf{M}$, a policy $\pi$, a deviation parameter $\delta$, and an STL property $\phi$, we say the system can \emph{tolerate} the deviation when the parametric form of $\mathbf{M}$ with parameter $\delta$
under the control of
$\pi$ satisfies the property, i.e., $\mathbf{M}^\delta || \pi \models \phi$.
\end{definition}

Then, the tolerance of a controller can be defined as all the possible deviations that the system can tolerate. Formally:
\begin{definition}\label{def:robustness}
For a system $\mathbf{M}$, a policy $\pi$, and an STL property $\phi$, the tolerance of the controller is defined as the \textbf{maximal} $\mathbf{\Delta} \subseteq \mathbb{R}^k$ s.t. $\forall \delta \in \mathbf{\Delta} : \mathbf{M}^{\delta}||\pi \models \phi$.
\end{definition}

In other words, the tolerance of a control policy $\pi$ is measured by the maximal parameter domain $\mathbf{\Delta}$ of a system where each deviated system $\mathbf{M}^\delta$ of it still satisfies the property under the control of $\pi$.

\subsection{Strict Evaluation of Tolerance}
In this work, we focus on a specific evaluation of tolerance. Specifically, Def. \ref{def:robust-system} and \ref{def:robustness} depend on the interpretation of $\mathbf{M}^{\delta} || \pi \models \phi$, i.e., a system satisfying a STL property; however, STL satisfaction is computed over a single trajectory. From the literature \cite{Corso2021-cps-survey}, one common evaluation criteria is that \emph{a system must not contain a trajectory that violates the STL property}. In other words, even in the worst-case scenario that is less likely to occur in a stochastic system, it should still guarantee the property. This interpretation enforces a strong guarantee of the system, and thus we call it the \emph{strict} satisfaction of STL in this work. Formally:
\begin{definition}\label{def:strict-stl-satisfaction}
A discrete-time stochastic system $\mathbf{M}$ \emph{strictly} satisfies an STL property $\phi$ under the control of a policy $\pi$ iff every controlled trajectory produces a non-negative STL robustness value, i.e.,
$\mathbf{M}||\pi \models \phi \Leftrightarrow \forall \sigma \in \mathcal{L}(\mathbf{M}||\pi) : \rho(\phi, \mathbf{s}_{\sigma}, 0) \geq 0$,
where $\mathbf{s}_{\sigma}$ is the signal of state values of trajectory $\sigma$.
\end{definition}

With this interpretation, we can then restate Def. \ref{def:robustness} as:
\begin{definition}\label{def:strict-robustness}
The tolerance of a policy $\pi$ that \emph{strictly} satisfies an STL property $\phi$ is the maximal $\mathbf{\Delta}$ s.t. $\forall \delta \in \mathbf{\Delta}, \sigma \in \mathcal{L}(\mathbf{M}^{\delta}||\pi) : \rho(\phi, \mathbf{s}_{\sigma}, 0) \geq 0$
\end{definition}
Although this definition delineates a strong tolerance guarantee, it can also be extended to more relaxed notions with probabilistic guarantees. In that case, other evaluation techniques for STL specification satisfaction such as \cite{fan2021statistical,Part-X,Lars2021-STL-risk} can be leveraged. We leave this as an extension of our work in the future.

\section{Tolerance Analysis}\label{sec:falsification}
\subsection{Tolerance Falsification}
According to Def. \ref{def:strict-robustness}, to compute the tolerance of a controller, we need to: (1) (formally) show that a stochastic system does not contain a trajectory that violates the STL property, and (2) compute the maximal parameter set $\mathbf{\Delta}$, which could be in any non-convex or even non-continuous shape, where all system instances $\mathbf{M}^\delta$ should satisfy step (1). This exhaustive computation is intractable due to the black-box RL controllers coupled with the stochasticity in system.


Therefore, in this work, instead of computing or approximating the tolerance $\mathbf{\Delta}$, we consider the problem of falsifying a given estimation of tolerance $\estrobustness$, i.e., finding a deviation $\delta \in \estrobustness$ that the system cannot tolerate for a given controller. More formally, we define:
\begin{problem}[Tolerance Falsification]\label{prob:robustness-falsify}
For a system $\mathbf{M}$, a policy $\pi$, and an STL property $\phi$, given a tolerance estimation $\estrobustness \subseteq \mathbb{R}^k$, the goal of a \emph{tolerance falsification problem} $\mathcal{F}(\mathbf{M}, \pi, \phi, \estrobustness)$ is to find a deviation $\delta \in \estrobustness$ s.t. $\exists \sigma \in \mathcal{L}(\mathbf{M}^\delta || \pi): \rho(\phi, \mathbf{s}_{\sigma}, 0) < 0$.
\end{problem}



\subsection{Minimum Tolerance Falsification}
Intuitively, a larger deviation (i.e., a deviation that is far away from the expected system parameter) would likely cause a larger deviation in the system behavior leading to a specification violation. However, controllers are generally not designed to handle arbitrarily large deviations in the first place,
and analyzing their performance in these situations offers limited insight to the designer. Moreover, if the designer decides to improve the tolerance of a controller (which is a costly endeavor), deviations closer to the nominal system are given high priority due to their higher likelihood of occurrence. In light of these practical design and deployment assumptions, we focus on the \emph{minimum} deviation problem.
\begin{problem}\label{prob:min-robustness-falsify}
Given a \emph{minimum} tolerance falsification problem $\mathcal{F}_{min}(\mathbf{M}, \pi, \phi, \estrobustness)$, let $\delta_0 \in \estrobustness$ be the zero-deviation point, the goal is to find a deviation $\delta \in \estrobustness$ s.t. $\mathbf{M}^\delta || \pi \nmodels \phi$ and $\delta$ minimizes a distance measure $\lVert \delta - \delta_0 \rVert_p$.
\end{problem}

\subsection{Falsification by Optimization}\label{sec:falsification-by-optimization}
Since the satisfaction of STL can be measured quantitatively, the tolerance falsification problem can be formulated as an optimization problem. Consider a real-valued system evaluation function $\Gamma(\mathbf{M}, \pi, \phi) \in \mathbb{R}$. We assume that if this function's value is negative, the controlled system violates the property, i.e., $$\Gamma(\mathbf{M}, \pi, \phi) < 0 \Leftrightarrow \mathbf{S}||\pi \nmodels \phi$$
and the smaller the value, the larger the degree of property violation.
Then, a tolerance falsification problem $\mathcal{F}(\mathbf{M}, \pi, \phi, \estrobustness)$ can be formulated as the following optimization problem:
\begin{align}\label{eq:any_deviation}
    \mathop{\arg\min}\limits_{\delta \in \estrobustness}~\Gamma(\mathbf{M}^{\delta}, \pi, \phi)
\end{align}
i.e., by finding a parameter $\delta \in \estrobustness$ that minimizes the evaluation function $\Gamma$ and observing this value can give information about system's property satisfaction. Concretely, if the minimum function value is negative, then the associated parameter $\delta$ indicates a deviation where the system violates the property $\phi$.
Specifically, in the case of strict evaluation of tolerance, the system evaluation function $\Gamma$ is defined as:
\begin{align}\label{eq:system_evaluation}
    \Gamma(\mathbf{M}, \pi, \phi) = \min \{ \rho(\phi, \mathbf{s}_{\sigma}, 0)~|~\sigma \in \mathcal{L}(\mathbf{M}||\pi) \}
\end{align}
Finally, we can formulate a minimum tolerance falsification problem $\mathcal{F}_{min}(\mathbf{M},$ $\pi, \phi, \estrobustness)$ as a constrained optimization problem:
\begin{align}\label{eq:min_deviation}
    \mathop{\arg\min}\limits_{\delta \in \estrobustness}~\lVert \delta - \delta_0 \rVert_p ~ s.t. ~ \Gamma(\mathbf{M}^{\delta}, \pi, \phi) < 0
\end{align}

Note that, Eq. \ref{eq:system_evaluation} is the typical formulation for solving a CPS falsification problem that intends to find a trajectory that violates an STL specification \cite{Corso2021-cps-survey}. Thus, the problem of finding \emph{any} tolerance violation (Eq. \ref{eq:any_deviation}) can be formulated as a min-min optimization problem which can be solved by existing CPS falsifiers such as Breach \cite{breach} and PsyTaLiRo \cite{S-TaliRo,Thibeault2021-psytaliro}.

However, the \emph{minimum} falsification problem (Eq. \ref{eq:min_deviation}) features multi-objective optimization or min-max optimization characteristics --- minimizing the deviation distance ($\lVert \delta - \delta_0 \rVert_p$) would likely cause a larger system evaluation value ($\Gamma$). Since these objectives are inherently conflicting, nuanced techniques are required to find solutions. Although, existing CPS falsifiers can be configured to represent this additional cost/objective function (either via specification modification or through explicit cost function definition), the underlying optimization techniques do not have a multi-layer setup to handle this off the shelf. Therefore, we present a novel two-layer search for solving the tolerance falsification problems, particularly effective in finding minimum violating deviations.

\section{Simulation-Based Tolerance Analysis Framework}\label{sec:framework}
\begin{figure}[!t]
\centering
\includegraphics[width=0.8\linewidth]{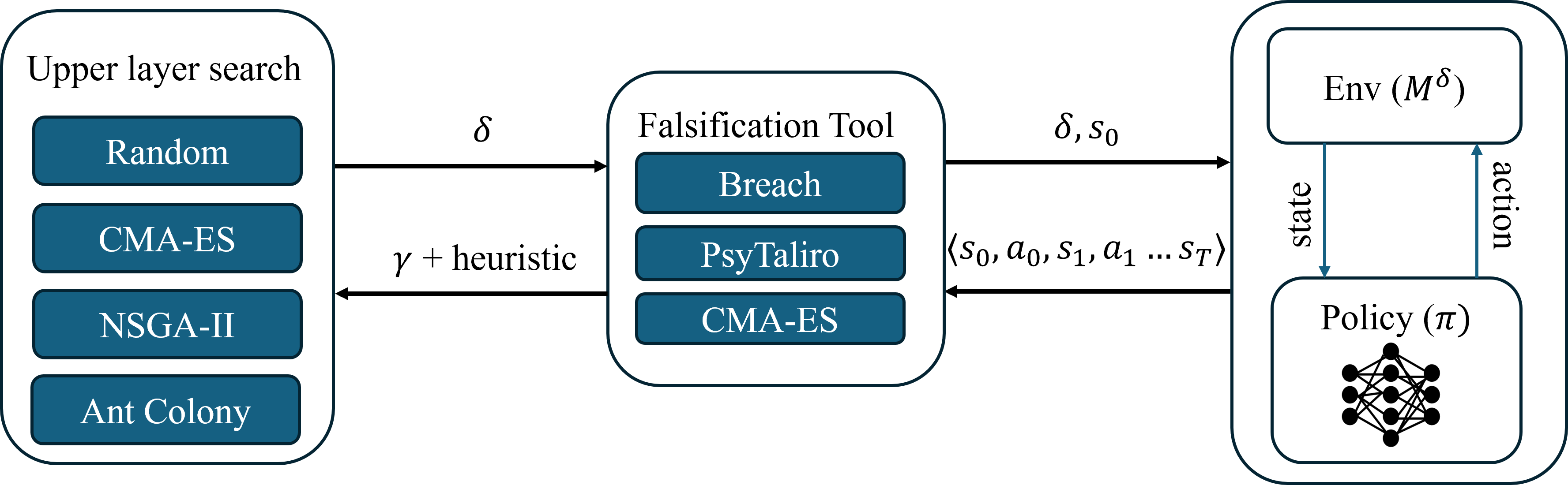}
\caption{Overview of the two-layer falsification algorithm.}
\label{fig:framework}
\vspace{-15pt}
\end{figure}

In this section, we outline our analysis framework to solve the tolerance falsification problems
for black-box CPS and RL controllers
(as shown in Figure \ref{fig:framework}).
We first explain our novel two layer falsification algorithm and then present a heuristic for more effective solving of the minimum falsification problem.

\subsection{A Two-Layer Falsification Algorithm}

Algorithm \ref{alg:falsification} presents our two-layer framework in details. Lines 3-13 indicate the upper-layer search. In each iteration, the upper-layer searches a set of deviation samples. For a deviation $\delta$, it instantiates a deviated system $\mathbf{M}^\delta$ (line 6), computes the system evaluation value $\gamma$ (line 7), and then computes the objective function value $v$ (line 8). The objective value indicates the quality of a deviation sample, e.g., whether it causes a violation of tolerance and has a small distance to the zero-deviation point. Finally, the objective values for this iteration is used to update the best result so far (line 11) and generates the next candidate solutions (line 12).
In particular, line 7 indicates the lower-layer task. It corresponds to the system evaluation function $\Gamma$ (which is the minimal STL robustness value according to Eq. \ref{eq:system_evaluation}).


Given the characteristics of our falsification problem, we propose this two-layer structure for multiple reasons: First, the separation of deviations and the lower-layer CPS falsification allows us to define richer evaluation metrics and heuristics that are solely relevant for deviation searching. These heuristics, if used in a single layer objective, would lead to an ill-posed optimization problem exacerbated by the highly non-convex landscapes of traditional CPS falsification.
Second, this separation of concerns allow us to find deviations closer to nominal points even for systems with high-dimensional state spaces, complex dynamics, and rugged robustness landscapes with multiple local minimas. In these settings, an one-layer search would converge to local solutions without exploring the search space extensively.
Finally, this two-layer structure provides us enough extensibility to:
\begin{itemize}
    \item Integrate many off-the-shelf optimization techniques for the upper-layer like we have for Uniform Random Sampling, CMA-ES \cite{cmaes}, NSGA-II \cite{nsga2}, and Extended Ant Colony \cite{gaco}.
    \item Integrate state-of-the-art CPS falsifiers (we integrated CMA-ES, Breach \cite{breach}, and PsyTaLiRo \cite{Thibeault2021-psytaliro}) and simulation platforms (we used OpenAI-Gym \cite{OpenAI-Gym}, PyBullet \cite{pybullet}, and Matlab Simulink). 
    
    \item Extend to other STL evaluation methods (function $\Gamma$), e.g., evaluation with probabilistic guarantees \cite{fan2021statistical,Part-X,Lars2021-STL-risk}, cumulative STL \cite{9029429}, or mean STL \cite{8814487}.
\end{itemize}

\begin{algorithm}[!t]
\SetKwFunction{Instantiate}{Instantiate}
\SetKwFunction{CPSFalsification}{CPSFalsification}
\SetKwFunction{NextCandidates}{NextCandidates}
\SetKwFunction{UpdateBest}{UpdateBest}
\SetKwInOut{Input}{Input}
\SetKwInOut{Output}{Output}
\Input{$\mathbf{M}, \pi, \phi, \estrobustness$, and objective function $f$}
\Output{violation $\delta_{best} \in \estrobustness$}

$\delta_{best} \gets nil$\;
$X \gets~\text{initial candidates from}~\estrobustness$ \;
\While{termination criteria = false}{
    $V \gets \langle \rangle$ \;
    \For{$\delta \in X$}{
        $M^\delta \gets$ \Instantiate{$\mathbf{M}^{\estrobustness}, \delta$} \;
        $\gamma \gets$ \CPSFalsification{$\mathbf{M}^\delta, \pi, \phi$} \;
        $v \gets f(\delta, \gamma)$ \tcp*{heuristic computation.}
        $V \gets V \frown \langle v \rangle$ \;
    }
    $\delta_{best} \gets$ \UpdateBest{X, V} \;
    $X \gets$ \NextCandidates{$f, X, V, \estrobustness$} \;
}

\caption{A Two-Layer Tolerance Falsification Algorithm}
\label{alg:falsification}
\end{algorithm}

\subsection{Heuristic for Efficient Minimum Tolerance Falsification}\label{sec:heuristic}
We present a novel heuristic for more effective discovery of minimum violating deviations.
Our heuristic is based on the known issues of RL policy overfitting. It has been highlighted in related literature that RL policies can overfit to the specific paramterized system used for training the policies and this dependence can reduce their applicability to real-world scenarios \cite{Peng2018-Sim-to-Real,Sadeghi2017cad2rl,Tobin2017domain}.
We exploit this over-fitting tendency to guide the search for $\delta$ that leads to a violation. Our heuristic is the cosine similarity between a deviated system's worst-case trajectory and a nominal system's worst-case trajectory. Formally:
\begin{equation*}
    dist(\delta) =\frac{\mathbf{Tr_\delta} \cdot \mathbf{Tr_{\delta_0}}}{\|\mathbf{Tr_\delta}\| \cdot \|\mathbf{Tr_{\delta_0}}\|}
\end{equation*}

Our intuition is that once a controller has been trained in a system parameterized by $\delta_0$, it overfits to that specific system. Then, when the controller is deployed in a deviated system, its worst-case trajectory will be similar to the nominal worst-case trajectory if
the distance between the two MDPs,  measured by the Euclidean distance between the parameters, is small. 
We measure the similarity between trajectories using cosine similarity. Thus, as the distance from the nominal MDP increases, the similarity score between the worst-case trajectories decreases.
This heuristic provides
more information about the search space: i.e. in the case there are two deviations where the robustness values are similar (which is possible due to the worst case semantics of STL robustness), cosine similarity can help in directing the search toward more violating directions.

\noindent{\textbf{Example.}} Concretely, we illustrate our heuristic's benefits through the CarRun system discussed in Section \ref{sec:motivation}.
First, the $\delta_0$ value (normative parameters) of the system is $\delta_0 = [20.0, 0.5]$, where the first one is the turn multiplier and the second one is the steering multiplier. Then, consider two concrete deviations $\delta_1 = [16.566, 0.409]$ and \ $\delta_2 = [15.136$, $0.447]$. The normalized $l\text{-2}$ distances of them to $\delta_0$ (i.e., $\lVert \delta - \delta_0 \rVert_2$) are 0.190 and 0.184, respectively. By solving the CPS falsification problems at these two deviations, their corresponding minimum STL robustness values are 0.130 and 0.125, respectively. That is, given a similar deviation distance, their worst-case robustness values are also close.
On the other hand, their corresponding worst-case trajectory similarity values are 0.900 and 0.995. Compared to the small difference in robustness values, this relatively big difference in similarity scores can better guide the upper-layer search to a violating deviation, i.e., the direction of $\delta_1$ might more likely lead to a violation and should be prioritized in the search.

\section{Evaluation}\label{sec:evaluation}

We implemented our proposed framework in a Python package\footnote{https://github.com/SteveZhangBit/STL-Robustness} and evaluate our technique through comprehensive experimentation. Our evaluation focuses on the minimum tolerance falsification problem.
Specifically, we measure our technique's effectiveness through three key metrics: (1) \emph{the number of violations found}, (2) \emph{the minimum distance of violations}, and (3) \emph{the average distance of violations}. Based on these metrics, we formulate the following research questions:
\begin{itemize}
    \item \textbf{RQ1}: Is our two-layer falsification framework more effective than leveraging an existing CPS falsifier?
    \item \textbf{RQ2}: Does our heuristic improve the effectiveness for finding minimum violating deviations, compared to off-the-self optimization algorithms?
\end{itemize}

Although existing CPS falsifiers \cite{breach,S-TaliRo,Thibeault2021-psytaliro} cannot directly solve our minimum tolerance falsification problem (Problem \ref{prob:min-robustness-falsify}), they allow customizing the objective function to optimize for both the deviation distance and STL robustness value to find minimum deviations. We call this technique \emph{one-layer search}. For RQ1, we benchmark against the one-layer search baseline for the minimum tolerance falsification problem. 
For RQ2, we evaluate whether our proposed heuristic described in Section \ref{sec:heuristic} further improves the effectiveness of our two-layer search, specifically the minimum distance.
\subsection{Experimental Setup and Implementational Details}

To answer these research questions, we first present a benchmark with systems and controllers trained to satisfy complex safety specifications.
The benchmark contains six systems with non linear dynamics adopted from OpenAI-Gym, PyBullet, and Matlab Simulink. We extend the interfaces of these systems so that users can configure their behavior for tolerance analysis by changing the system parameters. 

Then, we solve the corresponding minimum tolerance falsification problems for these problems. For each problem, we conduct the following experiments:
\begin{itemize}
    \item One-layer search leveraging an existing CPS falsifier by modifying the objective function to factor in the deviation distance and STL robustness value,
    \item Two-layer search with CMA-ES for both the upper and lower layers,
    \item Two-layer search with CMA-ES+Heuristic for the upper layer and CMA-ES for the lower layer.
\end{itemize}
Specifically, for the one-layer search, we employ the state-of-the-art CPS falsifiers, Breach \cite{breach} \edit{w/ CMA-ES} for Matlab systems and PsyTaLiRo \cite{Thibeault2021-psytaliro} \edit{w/ DualAnnealing} for Python systems by extending their default objective functions. For the two-layer search, due to the complexity of the CPS and the non-convex nature of STL robustness, the upper-layer optimization is also non-convex and has multiple local minima. Additionally, we assume black-box systems and controllers. Thus, due to these two considerations, we made the decision to adopt derivative-free evolutionary algorithms. Specifically, we primarily utilized CMA-ES as the upper-layer algorithm because it is widely used for black-box optimization and in our preliminary experiments outperformed other evolutionary methods. However, other algorithms can also be integrated. Furthermore, we also use CMA-ES for the lower-layer search as it is a widely used in CPS falsification tools \cite{Corso2021-cps-survey,breach} and works competitively for both Python and Matlab environments. 
Finally, we implement our heuristic and use it alongside the evaluation function for the upper-layer search.

Each problem was run three times on a Linux machine with a 3.6GHz CPU and 24GB memory. 
For fair evaluation, we set the budget in terms of the number of interactions with the simulator for all our techniques. Specifically, for one run, the budget for the one-layer search is 10,000 simulations; and the budget for the two-layer search is 100 for the upper-layer and 100 for the lower-layer falsification.


\subsection{Results}

\begin{table}[!t]
\centering
\renewcommand\arraystretch{1.3}
\caption{Minimum tolerance falsification results.
}
\label{tab:eval-results}
\resizebox{0.9\linewidth}{!}{
\begin{tabular}{lrrrrrrrrr}
\toprule
 & \multicolumn{3}{c}{\textbf{One-layer search}} & \multicolumn{3}{c}{\textbf{CMA-ES}} & \multicolumn{3}{c}{\textbf{CMA-ES w/ Heuristic}} \\ \cmidrule(lr){2-4} \cmidrule(lr){5-7} \cmidrule(lr){8-10}
 & \multicolumn{1}{r}{Viol.} & \multicolumn{1}{r}{Min Dst.} & \multicolumn{1}{r}{Avg. Dst.} & \multicolumn{1}{r}{Viol.} & \multicolumn{1}{r}{Min Dst.} & \multicolumn{1}{r}{Avg. Dst.} & \multicolumn{1}{r}{Viol.} & \multicolumn{1}{r}{Min Dst.} & \multicolumn{1}{r}{Avg. Dst.} \\ \midrule
Cartpole & \textbf{90} & 0.300 & \textbf{0.399} ~& 69 & 0.285 & 0.449~ & 79 & \textbf{0.256} & 0.417 \\
LunarLander & - & - & - ~& 74 & 0.026 & \textbf{0.222} ~& \textbf{84} & \textbf{0.020} & 0.293 \\
CarCircle & 11 & 0.143 & 0.255 ~& 22 & 0.102 & \textbf{0.219}~ & \textbf{57} & \textbf{0.068} & 0.454 \\
CarRun & 25 & 0.191 & \textbf{0.249} ~& 68 & 0.161 & 0.449~ & \textbf{109} & \textbf{0.156} & 0.399 \\
ACC & N/A & N/A & N/A~ & 43 & \textbf{0.110} & \textbf{0.323}~ & \textbf{110} & 0.138 & 0.415 \\
WTK & \textbf{300} & 0.299 & \textbf{0.443} ~& 54 & \textbf{0.296} & 0.454~ & 45 & 0.319 & 0.533 \\
\bottomrule
\end{tabular}
}
\vspace{-15pt}
\end{table}

Table \ref{tab:eval-results} summarizes the results for solving the minimum tolerance falsification problems. 
The \emph{Viol.} column shows the number of violations found in total from the three runs.
The \emph{Min Dst.} and \emph{Avg. Dst.} columns show the minimum and average normalized $l\text{-2}$ distance to the zero-deviation point (i.e., $\lVert \delta - \delta_0 \rVert_2$) of the found violations, respectively. The performance of our approach heavily depends on the underlying simulation time of a system that vastly outweighs the overhead added by our evolutionary search algorithms. Thus, we share comparable performance, measured by total run time, as tools like Breach and PsyTaLiRo given the same budget of simulation calls.

\begin{figure}[!t]
    \centering
    \includegraphics[width=0.9\textwidth]{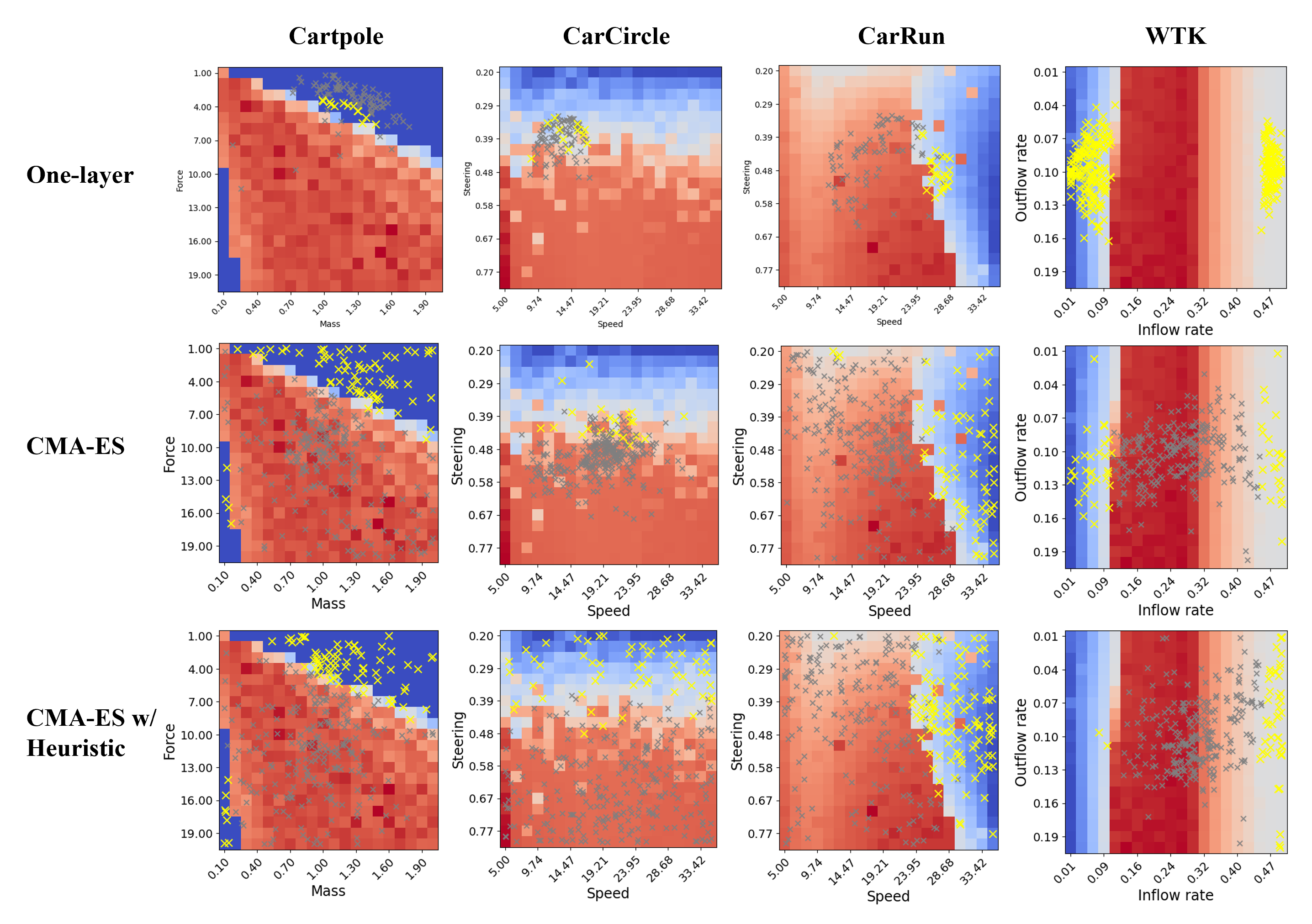}
    \caption{\footnotesize{Search spaces, deviation samples and violations processed by each algorithm.
    In each graph, the axes indicate the parameter domains. A red cell indicates a positive STL robustness value and a blue cell  a negative value. A grey cross indicates a deviation sample that is not falsified in the given budget; a yellow cross indicates a violation.
    }}
    \label{fig:heatmaps}
    \vspace{-15pt}
\end{figure}

In addition, to qualitatively exhibit our approach's effectiveness in finding deviations, we visualize the search space landscape for different problems in heat maps. Each heat map is generated by slicing the space (i.e., the estimated domain of system parameters) into a $20\times20$ grid and using a CPS falsifier to find the minimum STL robustness value for each grid cell. However, this processing is only done for visualization purposes and is not used in any of the algorithms.
This brute force sampling requires far more resources than our falsification approach. 
Finally, we draw the deviation samples and violations from our analysis on the heat maps. 
The final results are illustrated visually in Figure \ref{fig:heatmaps}.


\noindent{\textbf{Answer to RQ1.}} As illustrated in the table, the one-layer search fails to find violations in the LunarLander problem, and it cannot represent the type of system parameters we need in the ACC problem (due to falsification tool implementation). On the other hand, our two-layer search with CMA-ES solves all the problems and finds smaller deviations than the one-layer search in all problems. Moreover, as can be observed in the heat maps, since the distance value is directly appended to the STL robustness value in the one-layer search, it fails to find small deviations that \textit{barely} violate the property because it would result in a larger objective value. Thus, it is hard for it to converge to the minimum violating deviations. On the other hand, our two-layer search can better converge to the boundary of safe and unsafe regions. However, it also causes it to find fewer violations because it searches for more samples in the safe region close to the boundary where violations can be rare.


\noindent{\textbf{Answer to RQ2.}} From the table, our two-layer search with CMA-ES+Heuristic finds smaller violating deviations than the original CMA-ES in 4/6 problems.
It also finds more violations in 5/6 problems. However, the average distances also increase in 4/6 problems due to more exploration of violations encouraged by our heuristic. Despite that, from the heat maps, our CMA-ES+Heuristic approach can still converge to small violating deviations on the safe and unsafe boundary while also finding more violations.
Our heuristic helps in guiding the search and provides additional information to the algorithm when STL robustness is not enough to provide directionality. Concretely, a small similarity value would likely lead to a violation (even when the robustness value is similar) and thus results in more violations found and faster convergence to a small violation.

\section{Related Work}\label{sec:related_work}
There exists similar CPS tolerance notions from a control theory perspective such as \cite{Saoud23-CPS-resilience,Fainekos09-LPV}. For example, Saoud et al. \cite{Saoud23-CPS-resilience} present a resilience notion of CPS based on LTL w.r.t. a real-valued disturbance space. Then, they present an optimization-based method to approximate the maximum set of disturbances that maintain a desired LTL property for linear control systems. These notions target traditional controllers with a white-box assumption of systems and controllers, whereas we employ a black-box assumption which is more practical regarding complex CPS and NN-based RL controllers. 


Falsification of CPS \cite{Corso2021-cps-survey} is a well-studied problem in the literature. The goal is to find counterexample trajectories that violate a STL property by mutating the initial states or system inputs. A related application is parameter synthesis \cite{bartocci2018specification-survey} that finds a set of system parameters where the system satisfies the property. It can be seen as a dual problem to the falsification problem. Tools like Breach \cite{breach} and PSY-TaLiRo \cite{Thibeault2021-psytaliro,S-TaliRo} support both types of analysis. However, our tolerance falsification problem can be seen as solving these two problems at the same time. Our upper-layer search aims to find system parameters that would lead to a violation of the system specification, and the lower-layer search aims to find initial states or system inputs that lead to a violating trajectory. Although our problem can be reduced to a CPS falsification problem with system parameters, it is not effective in solving our minimum tolerance falsification problem compared to our two-layer structure as illustrated by our experimental results.

VerifAI \cite{VerifAI,Sanjit-compositional} applies a similar idea to us where they consider \emph{abstract features} for a ML model that can lead to a property violation of a CPS. Different from us, they assume a CPS with a ML perception model (such as object detection) connecting to a traditional controller, and the abstract features are environmental parameters that would affect the performance of the ML model (e.g., brightness). In other words, they focus on deviations that affect the ML model whereas our deviation notion is more general that includes any external or internal deviation or sensor error which changes the system dynamics.


Robust RL studies the problem to improve the performance of RL controllers in the presence of uncertainties
\cite{Moos-RobustRL, xu2022trustworthy}. A similar research topic is domain randomization \cite{Peng2018-Sim-to-Real,Sadeghi2017cad2rl,Tobin2017domain} that create various systems with randomized parameters leading to changed system dynamics and then train a controller that works across these systems. However, our work is different in that: (1) we focus on tolerance evaluation whereas they focus more on training; and (2) we focus on system specifications and specify them in STL properties, while they rely on rewards where maximizing the reward does not necessarily guarantee certain system specification.

\todo{Add related work from Andre Platzer}

\section{Conclusion}\label{sec:conclusion}
In this paper, we have introduced a specification-based tolerance definition for CPS.
This definition yields a new type of analysis problem, called  \emph{tolerance falsification}, where the goal is to find small changes to the system dynamics that result in a violation of a given STL specification. We have also presented a novel optimization-based approach to solve the problem and evaluated the effectiveness of it over our proposed CPS tolerance analysis benchmark.

Since our analysis framework is extensible, as part of future work, we plan to explore and integrate other types of evaluation functions $\Gamma$ (e.g., evaluation with probabilistic guarantees \cite{Part-X,Lars2021-STL-risk,fan2021statistical}), different semantics of STL robustness (e.g., cumulative robustness\cite{9029429}), or leveraging decomposition of STL for more effective falsification of complex specifications \cite{kapoor2024safe}. 
Moreover, we currently use $l\text{-2}$ norm to compute the deviation distances. In the future, we also plan to explore other distance notions such as Wasserstein Distance \cite{Lecarpentier2019-wasserstein,Abdullah2019-wasserstein,Yang2017-wasserstein}, which computes distribution distance between system dynamics.

\bibliographystyle{IEEEtran}
\bibliography{IEEEabrv,ref}

\begin{thebibliography}{10}
\providecommand{\url}[1]{#1}
\csname url@samestyle\endcsname
\providecommand{\newblock}{\relax}
\providecommand{\bibinfo}[2]{#2}
\providecommand{\BIBentrySTDinterwordspacing}{\spaceskip=0pt\relax}
\providecommand{\BIBentryALTinterwordstretchfactor}{4}
\providecommand{\BIBentryALTinterwordspacing}{\spaceskip=\fontdimen2\font plus
\BIBentryALTinterwordstretchfactor\fontdimen3\font minus \fontdimen4\font\relax}
\providecommand{\BIBforeignlanguage}[2]{{%
\expandafter\ifx\csname l@#1\endcsname\relax
\typeout{** WARNING: IEEEtran.bst: No hyphenation pattern has been}%
\typeout{** loaded for the language `#1'. Using the pattern for}%
\typeout{** the default language instead.}%
\else
\language=\csname l@#1\endcsname
\fi
#2}}
\providecommand{\BIBdecl}{\relax}
\BIBdecl

\bibitem{Collins2018QuantifyingTR}
\BIBentryALTinterwordspacing
J.~J. Collins, D.~Howard, and J.~Leitner, ``Quantifying the reality gap in robotic manipulation tasks,'' \emph{2019 International Conference on Robotics and Automation (ICRA)}, pp. 6706--6712, 2018. [Online]. Available: \url{https://api.semanticscholar.org/CorpusID:53208962}
\BIBentrySTDinterwordspacing

\bibitem{Moos-RobustRL}
\BIBentryALTinterwordspacing
J.~Moos, K.~Hansel, H.~Abdulsamad, S.~Stark, D.~Clever, and J.~Peters, ``Robust reinforcement learning: A review of foundations and recent advances,'' \emph{Machine Learning and Knowledge Extraction}, vol.~4, no.~1, pp. 276--315, 2022. [Online]. Available: \url{https://www.mdpi.com/2504-4990/4/1/13}
\BIBentrySTDinterwordspacing

\bibitem{xu2022trustworthy}
M.~Xu, Z.~Liu, P.~Huang, W.~Ding, Z.~Cen, B.~Li, and D.~Zhao, ``Trustworthy reinforcement learning against intrinsic vulnerabilities: Robustness, safety, and generalizability,'' 2022.

\bibitem{Peng2018-Sim-to-Real}
X.~B. Peng, M.~Andrychowicz, W.~Zaremba, and P.~Abbeel, ``Sim-to-real transfer of robotic control with dynamics randomization,'' in \emph{2018 IEEE International Conference on Robotics and Automation (ICRA)}, 2018, pp. 3803--3810.

\bibitem{Sadeghi2017cad2rl}
F.~Sadeghi and S.~Levine, ``Cad2rl: Real single-image flight without a single real image,'' 2017.

\bibitem{Tobin2017domain}
J.~Tobin, R.~Fong, A.~Ray, J.~Schneider, W.~Zaremba, and P.~Abbeel, ``Domain randomization for transferring deep neural networks from simulation to the real world,'' 2017.

\bibitem{10.5555/645528.657613}
A.~Y. Ng, D.~Harada, and S.~J. Russell, ``Policy invariance under reward transformations: Theory and application to reward shaping,'' in \emph{Proceedings of the Sixteenth International Conference on Machine Learning}, ser. ICML '99.\hskip 1em plus 0.5em minus 0.4em\relax San Francisco, CA, USA: Morgan Kaufmann Publishers Inc., 1999, p. 278–287.

\bibitem{Booth_Knox_Shah_Niekum_Stone_Allievi_2023}
\BIBentryALTinterwordspacing
S.~Booth, W.~B. Knox, J.~Shah, S.~Niekum, P.~Stone, and A.~Allievi, ``The perils of trial-and-error reward design: Misdesign through overfitting and invalid task specifications,'' \emph{Proceedings of the AAAI Conference on Artificial Intelligence}, vol.~37, no.~5, pp. 5920--5929, Jun. 2023. [Online]. Available: \url{https://ojs.aaai.org/index.php/AAAI/article/view/25733}
\BIBentrySTDinterwordspacing

\bibitem{stl}
A.~Donz{\'e} and O.~Maler, ``Robust satisfaction of temporal logic over real-valued signals,'' in \emph{Formal Modeling and Analysis of Timed Systems}, K.~Chatterjee and T.~A. Henzinger, Eds.\hskip 1em plus 0.5em minus 0.4em\relax Berlin, Heidelberg: Springer Berlin Heidelberg, 2010, pp. 92--106.

\bibitem{handbook-probabilistic}
C.~Baier, L.~de~Alfaro, V.~Forejt, and M.~Kwiatkowska, \emph{Model Checking Probabilistic Systems}.\hskip 1em plus 0.5em minus 0.4em\relax Cham: Springer International Publishing, 2018, pp. 963--999.

\bibitem{sutton2018reinforcement}
R.~S. Sutton and A.~G. Barto, \emph{Reinforcement learning: An introduction}.\hskip 1em plus 0.5em minus 0.4em\relax MIT press, 2018.

\bibitem{10.5555/2789272.2886795}
J.~Garc\'{\i}a and F.~Fern\'{a}ndez, ``A comprehensive survey on safe reinforcement learning,'' \emph{J. Mach. Learn. Res.}, vol.~16, no.~1, p. 1437–1480, jan 2015.

\bibitem{Gu2022ARO}
\BIBentryALTinterwordspacing
S.~Gu, L.~Yang, Y.~Du, G.~Chen, F.~Walter, J.~Wang, Y.~Yang, and A.~Knoll, ``A review of safe reinforcement learning: Methods, theory and applications,'' \emph{ArXiv}, vol. abs/2205.10330, 2022. [Online]. Available: \url{https://api.semanticscholar.org/CorpusID:248965265}
\BIBentrySTDinterwordspacing

\bibitem{yu2018policy}
\BIBentryALTinterwordspacing
W.~Yu, C.~K. Liu, and G.~Turk, ``Policy transfer with strategy optimization,'' in \emph{International Conference on Learning Representations}, 2019. [Online]. Available: \url{https://openreview.net/forum?id=H1g6osRcFQ}
\BIBentrySTDinterwordspacing

\bibitem{Bhattacharyya:1995}
S.~P. Bhattacharyya, H.~Chapellat, and L.~H. Keel, \emph{Robust Control: The Parametric Approach}, 1st~ed.\hskip 1em plus 0.5em minus 0.4em\relax USA: Prentice Hall PTR, 1995.

\bibitem{weinmann:2012uncertain}
A.~Weinmann, \emph{Uncertain models and robust control}.\hskip 1em plus 0.5em minus 0.4em\relax Springer Science \& Business Media, 2012.

\bibitem{Corso2021-cps-survey}
A.~Corso, R.~Moss, M.~Koren, R.~Lee, and M.~Kochenderfer, ``A survey of algorithms for black-box safety validation of cyber-physical systems,'' \emph{Journal of Artificial Intelligence Research}, vol.~72, pp. 377--428, 2021.

\bibitem{fan2021statistical}
C.~Fan, X.~Qin, Y.~Xia, A.~Zutshi, and J.~Deshmukh, ``Statistical verification of autonomous systems using surrogate models and conformal inference,'' 2021.

\bibitem{Part-X}
G.~Pedrielli, T.~Khandait, Y.~Cao, Q.~Thibeault, H.~Huang, M.~Castillo-Effen, and G.~Fainekos, ``Part-x: A family of stochastic algorithms for search-based test generation with probabilistic guarantees,'' \emph{IEEE Transactions on Automation Science and Engineering}, pp. 1--22, 2023.

\bibitem{Lars2021-STL-risk}
L.~Lindemann, N.~Matni, and G.~J. Pappas, ``Stl robustness risk over discrete-time stochastic processes,'' in \emph{2021 60th IEEE Conference on Decision and Control (CDC)}, 2021, pp. 1329--1335.

\bibitem{breach}
A.~Donz{\'e}, ``Breach, a toolbox for verification and parameter synthesis of hybrid systems,'' in \emph{Computer Aided Verification}, T.~Touili, B.~Cook, and P.~Jackson, Eds.\hskip 1em plus 0.5em minus 0.4em\relax Berlin, Heidelberg: Springer Berlin Heidelberg, 2010, pp. 167--170.

\bibitem{S-TaliRo}
Y.~Annpureddy, C.~Liu, G.~Fainekos, and S.~Sankaranarayanan, ``S-taliro: A tool for temporal logic falsification for hybrid systems,'' in \emph{Tools and Algorithms for the Construction and Analysis of Systems}, P.~A. Abdulla and K.~R.~M. Leino, Eds.\hskip 1em plus 0.5em minus 0.4em\relax Berlin, Heidelberg: Springer Berlin Heidelberg, 2011, pp. 254--257.

\bibitem{Thibeault2021-psytaliro}
Q.~Thibeault, J.~Anderson, A.~Chandratre, G.~Pedrielli, and G.~Fainekos, ``Psy-taliro: A python toolbox for search-based test generation for cyber-physical systems,'' 2021.

\bibitem{cmaes}
N.~Hansen and A.~Ostermeier, ``Adapting arbitrary normal mutation distributions in evolution strategies: the covariance matrix adaptation,'' in \emph{Proceedings of IEEE International Conference on Evolutionary Computation}, 1996, pp. 312--317.

\bibitem{nsga2}
K.~Deb, A.~Pratap, S.~Agarwal, and T.~Meyarivan, ``A fast and elitist multiobjective genetic algorithm: Nsga-ii,'' \emph{IEEE Transactions on Evolutionary Computation}, vol.~6, no.~2, pp. 182--197, 2002.

\bibitem{gaco}
\BIBentryALTinterwordspacing
M.~Schlüter, J.~A. Egea, and J.~R. Banga, ``Extended ant colony optimization for non-convex mixed integer nonlinear programming,'' \emph{Computers \& Operations Research}, vol.~36, no.~7, pp. 2217--2229, 2009. [Online]. Available: \url{https://www.sciencedirect.com/science/article/pii/S0305054808001524}
\BIBentrySTDinterwordspacing

\bibitem{OpenAI-Gym}
G.~Brockman, V.~Cheung, L.~Pettersson, J.~Schneider, J.~Schulman, J.~Tang, and W.~Zaremba, ``Openai gym,'' 2016.

\bibitem{pybullet}
E.~Coumans and Y.~Bai, ``Pybullet, a python module for physics simulation for games, robotics and machine learning,'' \url{http://pybullet.org}, 2016.

\bibitem{9029429}
I.~Haghighi, N.~Mehdipour, E.~Bartocci, and C.~Belta, ``Control from signal temporal logic specifications with smooth cumulative quantitative semantics,'' in \emph{2019 IEEE 58th Conference on Decision and Control (CDC)}, 2019, pp. 4361--4366.

\bibitem{8814487}
N.~Mehdipour, C.-I. Vasile, and C.~Belta, ``Arithmetic-geometric mean robustness for control from signal temporal logic specifications,'' in \emph{2019 American Control Conference (ACC)}, 2019, pp. 1690--1695.

\bibitem{Saoud23-CPS-resilience}
A.~Saoud, P.~Jagtap, and S.~Soudjani, ``Temporal logic resilience for cyber-physical systems,'' in \emph{2023 62nd IEEE Conference on Decision and Control (CDC)}, 2023, pp. 2066--2071.

\bibitem{Fainekos09-LPV}
G.~E. Fainekos and G.~J. Pappas, ``Mtl robust testing and verification for lpv systems,'' in \emph{2009 American Control Conference}, 2009, pp. 3748--3753.

\bibitem{bartocci2018specification-survey}
E.~Bartocci, J.~Deshmukh, A.~Donz{\'e}, G.~Fainekos, O.~Maler, D.~Ni{\v{c}}kovi{\'c}, and S.~Sankaranarayanan, ``Specification-based monitoring of cyber-physical systems: a survey on theory, tools and applications,'' \emph{Lectures on Runtime Verification: Introductory and Advanced Topics}, pp. 135--175, 2018.

\bibitem{VerifAI}
T.~Dreossi, D.~J. Fremont, S.~Ghosh, E.~Kim, H.~Ravanbakhsh, M.~Vazquez-Chanlatte, and S.~A. Seshia, ``Verifai: A toolkit for the formal design and analysis of artificial intelligence-based systems,'' in \emph{Computer Aided Verification}, I.~Dillig and S.~Tasiran, Eds.\hskip 1em plus 0.5em minus 0.4em\relax Cham: Springer International Publishing, 2019, pp. 432--442.

\bibitem{Sanjit-compositional}
T.~Dreossi, A.~Donz{\'e}, and S.~A. Seshia, ``Compositional falsification of cyber-physical systems with machine learning components,'' \emph{Journal of Automated Reasoning}, vol.~63, pp. 1031--1053, 2019.

\bibitem{kapoor2024safe}
P.~Kapoor, E.~Kang, and R.~Meira-Goes, ``Safe planning through incremental decomposition of signal temporal logic specifications,'' \emph{arXiv preprint arXiv:2403.10554}, 2024.

\bibitem{Lecarpentier2019-wasserstein}
E.~Lecarpentier and E.~Rachelson, ``Non-stationary markov decision processes, a worst-case approach using model-based reinforcement learning,'' in \emph{Advances in Neural Information Processing Systems}, H.~Wallach, H.~Larochelle, A.~Beygelzimer, F.~d\textquotesingle Alch\'{e}-Buc, E.~Fox, and R.~Garnett, Eds., vol.~32.\hskip 1em plus 0.5em minus 0.4em\relax Curran Associates, Inc., 2019.

\bibitem{Abdullah2019-wasserstein}
M.~A. Abdullah, H.~Ren, H.~B. Ammar, V.~Milenkovic, R.~Luo, M.~Zhang, and J.~Wang, ``Wasserstein robust reinforcement learning,'' 2019.

\bibitem{Yang2017-wasserstein}
I.~Yang, ``A convex optimization approach to distributionally robust markov decision processes with wasserstein distance,'' \emph{IEEE Control Systems Letters}, vol.~1, no.~1, pp. 164--169, 2017.

\bibitem{Mnih2013-DQN}
V.~Mnih, K.~Kavukcuoglu, D.~Silver, A.~Graves, I.~Antonoglou, D.~Wierstra, and M.~Riedmiller, ``Playing atari with deep reinforcement learning,'' 2013.

\bibitem{Schulman2017-PPO}
J.~Schulman, F.~Wolski, P.~Dhariwal, A.~Radford, and O.~Klimov, ``Proximal policy optimization algorithms,'' 2017.

\bibitem{Gronauer2022BulletSafetyGym}
S.~Gronauer, ``Bullet-safety-gym: A framework for constrained reinforcement learning,'' mediaTUM, Tech. Rep., 2022.

\bibitem{Liu2023robustness}
Z.~Liu, Z.~Guo, Z.~Cen, H.~Zhang, J.~Tan, B.~Li, and D.~Zhao, ``On the robustness of safe reinforcement learning under observational perturbations,'' 2023.

\bibitem{Haarnoja2018-SAC}
T.~Haarnoja, A.~Zhou, P.~Abbeel, and S.~Levine, ``Soft actor-critic: Off-policy maximum entropy deep reinforcement learning with a stochastic actor,'' 2018.

\bibitem{AI-CPS}
\BIBentryALTinterwordspacing
J.~Song, D.~Lyu, Z.~Zhang, Z.~Wang, T.~Zhang, and L.~Ma, ``When cyber-physical systems meet ai: A benchmark, an evaluation, and a way forward,'' in \emph{Proceedings of the 44th International Conference on Software Engineering: Software Engineering in Practice}, ser. ICSE-SEIP '22.\hskip 1em plus 0.5em minus 0.4em\relax New York, NY, USA: Association for Computing Machinery, 2022, p. 343–352. [Online]. Available: \url{https://doi.org/10.1145/3510457.3513049}
\BIBentrySTDinterwordspacing

\bibitem{Fujimoto2018-TD3}
\BIBentryALTinterwordspacing
S.~Fujimoto, H.~van Hoof, and D.~Meger, ``Addressing function approximation error in actor-critic methods,'' in \emph{Proceedings of the 35th International Conference on Machine Learning}, ser. Proceedings of Machine Learning Research, J.~Dy and A.~Krause, Eds., vol.~80.\hskip 1em plus 0.5em minus 0.4em\relax PMLR, 10--15 Jul 2018, pp. 1587--1596. [Online]. Available: \url{https://proceedings.mlr.press/v80/fujimoto18a.html}
\BIBentrySTDinterwordspacing

\end{thebibliography}

\clearpage

\section*{Appendix}

\subsection*{Benchmark Problem Descriptions}
The following sections describe the details about the systems of our CPS tolerance evaluation benchmark.

\subsubsection{Cart-Pole}
The Cart-Pole problem is described in Section \ref{sec:motivation}. In our experiments, we synthesize a PID and a DQN \cite{Mnih2013-DQN} controller for it; and we define four deviation dimensions, the \emph{Mass of the cart}, the \emph{Mass of the pole}, the \emph{Length of the pole}, and the \emph{Force} when pushing the cart.

\subsubsection{Lunar-Lander}
The Lunar-Lander system\footnote{https://www.gymlibrary.dev/environments/box2d/lunar\_lander/} where the goal is to control an aircraft to safely land on the surface of a planet (within the flagged area). It can fire the main engine (on the bottom) and the left/right engines to control the pose of the aircraft. The safety property defines \emph{1) the rotation of the aircraft should be within a value $\theta$ (e.g., not parallel to the ground or upside down), and 2) it should be close to the landing target as the height decreasing.} In our experiments, we develop a LQR and a PPO \cite{Schulman2017-PPO} controller for it; and we define three deviation dimensions, the \emph{Wind} that can change the x-y position of the aircraft, the \emph{Turbulence} (rotational wind) that can change the rotation of it, and the \emph{Gravity}.

\subsubsection{Car-Circle}
The Car-Circle system where the task is to control a car to move along the circumference of the blue circle \cite{Gronauer2022BulletSafetyGym}. There are ``walls'' on the two sides and the safety property defines \emph{the car should not move across the walls}. In our experiments, we leverage a PPO variation for it from \cite{Liu2023robustness} (which is more robust than standard PPO in the context of robust RL); and we define three types of deviations, the \emph{Force} that moves the car, the \emph{Speed Multiplier} and the \emph{Steering Multiplier} that affect the sensitivity of the forward velocity and the angular velocity response to the force, respectively.

\subsubsection{Car-Run}
The Car-Run system where the task is to control a car move along the track without hitting the walls on the two sides \cite{Gronauer2022BulletSafetyGym}. That is, the safety property defines \emph{the car should not move across the walls}. In our experiments, similar to the Car-Circle system, we also leverage a PPO variation from \cite{Liu2023robustness} and consider the \emph{Speed Multiplier} and the \emph{Steering Multiplier} deviation types.

\subsubsection{Adaptive Cruise Control}
A vehicle equipped with adaptive cruise control (ACC)\footnote{https://www.mathworks.com/help/mpc/ug/adaptive-cruise-control-using-model-predictive-controller.html} has a sensor that measures the distance to the preceding vehicle in the same line. The control goal is to: 1) control the speed of the vehicle to reach the driver-set velocity, and 2) maintain a safe distance to the leading vehicle. Therefore, we have the safety property that \emph{the relative distance between the ego vehicle and the leading vehicle should always be greater than a safe distance}. In our experiments, we adopt an MPC controller from Matlab and a SAC \cite{Haarnoja2018-SAC} controller from Jiayang Song et. al \cite{AI-CPS}. We define three types of deviations, the \emph{Mass} of the vehicle, and the \emph{min} and \emph{max acceleration of the leading vehicle}, changing which can mimic a more progressive or conservative leading vehicle that changes its speed more abruptly or slowly.

\subsubsection{Water Tank}
A water tank (WTK) system is a container with a controller controlling the inflow and outflow of water, widely used in industry domains like the chemical industry.\footnote{https://www.mathworks.com/help/slcontrol/gs/watertank-simulink-model.html} The safety property is defined such that \emph{the error between the actual water level and the desired water level should always be below a threshold}. We adopt a PID controller from Matlab and a TD3 \cite{Fujimoto2018-TD3} controller from \cite{AI-CPS}. We define two types of deviations, \emph{the water flow rate into the tank} and \emph{the water flow rate out of the tank}, which affect how fast the water volume would change.

\end{document}